\documentclass[twocolumn,amsmath,amssymb,pra,superscriptaddress,longbibliography]{revtex4-1}

\setcitestyle{super}

\usepackage[final,letterspace=-40]{microtype}
\usepackage[colorlinks=true, linkcolor=blue, urlcolor=blue,  citecolor=blue, anchorcolor=blue]{hyperref}

\usepackage{mathpazo} 

\usepackage{graphicx}
\usepackage{dcolumn}
\usepackage{bm}
\usepackage{helvet}

\usepackage{caption}
\DeclareCaptionLabelFormat{plain}{Figure #2 $\boldsymbol |$ }

\makeatletter
\renewcommand*{\@fnsymbol}[1]{\ensuremath{\ifcase#1\or *\or *,\dagger\or \ddagger\or
    \mathsection\or \mathparagraph\or \|\or **\or \dagger\dagger
    \or \ddagger\ddagger \else\@ctrerr\fi}}
\makeatother

\begin{document}

\title{Demonstration of an optical-coherence converter}

\author{Chukwuemeka O. Okoro}
\thanks{These authors contributed equally.}
\affiliation{Department of Electrical and Computer Engineering, University of Illinois at Urbana–Champaign, Urbana, Illinois 61801, USA}
\author{H. Esat Kondakci}
\email{esat@creol.ucf.edu}
\author{Ayman F. Abouraddy}
\affiliation{CREOL, The College of Optics \& Photonics, University of Central Florida, Orlando, Florida 32816, USA}
\author{Kimani C. Toussaint, Jr.}
\affiliation{Department of Mechanical Science and Engineering, University of Illinois at Urbana–Champaign, Urbana, Illinois 61801, USA}

\begin{abstract} \noindent 	
Studying the coherence of an optical field is typically compartmentalized with respect to its different optical degrees of freedom (DoFs) -- spatial, temporal, and polarization. Although this traditional approach succeeds when the DoFs are uncoupled, it fails at capturing key features of the field's coherence if the DOFs are indeed correlated -- a situation that arises often. By viewing coherence as a `resource' that can be shared among the DoFs, it becomes possible to convert the entropy associated with the fluctuations in one DoF to another DoF that is initially fluctuation-free. Here, we verify experimentally that coherence can indeed be reversibly exchanged -- without loss of energy -- between polarization and the spatial DoF of a partially coherent field. Starting from a linearly polarized spatially incoherent field -- one that produces no spatial interference fringes -- we obtain a spatially coherent field that is unpolarized. By reallocating the entropy to polarization, the field becomes invariant with regards to the action of a polarization scrambler, thus suggesting a strategy for avoiding the deleterious effects of a randomizing system on a DoF of the optical field.
\end{abstract}

\small
\maketitle


\noindent 
Optical coherence is evaluated by assessing the correlations between field fluctuations at different points in space and time \cite{Mandel95Book}. When multiple degrees of freedom (DoFs) of an optical field -- spatial, temporal, and polarization -- are relevant, the coherence of each DoF is typically studied separately. For example, spatial coherence is evaluated via double-slit interference \cite{Zernike38P}, temporal coherence through two-path (e.g., Michelson) interference \cite{Born99Book}, and polarization coherence by measuring the Stokes parameters \cite{Brosseau98Book}. Although this traditional approach succeeds when the DoFs are uncoupled, it fails at capturing key features of the field coherence if they are correlated \cite{Qian11OL,Kagalwala13NP}.

%

Here we show that coherence can be viewed as a `resource' that can be reversibly converted from one DoF of the field to another. We demonstrate experimentally the reversible and energy-conserving (unitary) conversion of coherence between the spatial and polarization DoFs of an optical field. Starting from a linearly polarized field having \textit{no spatial coherence} (a complete lack of double-slit interference visibility), we convert the field without filtering or loss of energy into one that displays spatial coherence (high-visibility interference fringes) -- but is \textit{unpolarized}. The optical arrangement we describe engenders an internal reorganization of the field energy that leads to a migration of the entropy associated with the statistical fluctuations from one DoF (spatial) to another (polarization). This coherency conversion is confirmed by measuring the full $4\times4$ coherency matrix that provides a complete description of two-point vector-field coherence via `optical coherency matrix tomography' (OCmT) \cite{Abouraddy14OL,Kagalwala15SR}. The tomographic measurement of coherence is carried out at different stages in the experimental setup to confirm the transformations involved in the coherence conversion process. As an application to highlight the usefulness of reallocating the entropy from one DoF of the field to another, we show that the field can be reconfigured to be invariant under the impact of a depolarizer or polarization scrambler that transforms any input polarization to unpolarized light. By transferring all the field entropy into polarization, the polarization scrambler cannot further increase the polarization entropy, which thus emerges unchanged. Since the coherence conversion procedure is reversible and no energy is lost, the field may be reversed to its original fully polarized configuration after traversing the polarization scrambler.

The paper first briefly reviews the matrix approach to quantifying optical coherence for a single DoF and multiple DoFs. Second, we describe the concept of an optical-coherence converter, a system that reversibly transforms coherence -- viewed as a resource -- from one DoF of a field to another without loss of energy. Starting from a field with one coherent DoF and another incoherent DoF, we reversibly convert the field such that the former DoF becomes incoherent and the latter coherent. Third, we present the experimental arrangement used in confirming these predictions and the measurement scheme to identify multi-DoF beam coherence. Finally, we demonstrate the invariance of a field with respect to a polarization scrambler before presenting our conclusions.

\section{Multi-DoF coherence}

\subsection{Polarization coherence and spatial coherence}

Partial polarization is described by a polarization coherency matrix $\mathbf{G}_{\mathrm{p}}=\left(\small{\begin{array}{cc}G^{\mathrm{HH}}&G^{\mathrm{HV}}\\G^{\mathrm{VH}}&G^{\mathrm{VV}}\end{array}}\right)$,
where $\mathrm{H}$ and $\mathrm{V}$ identify the horizontal and vertical polarization components, respectively, $G^{ij}=\langle E^{i}(E^{j})^{*}\rangle$, $i,j=\mathrm{H},\mathrm{V}$, and $\langle\cdot\rangle$ is an ensemble average \cite{Wolf07Book}. Here $\mathbf{G}_{\mathrm{p}}$ is Hermitian ($\mathbf{G}_{\mathrm{p}}^{\dag}=\mathbf{G}_{\mathrm{p}}$), semi-positive, normalized such that $G^{\mathrm{HH}}+G^{\mathrm{VV}}=1$, $G^{\mathrm{HH}}$ and $G^{\mathrm{VV}}$ are the contributions of H and V to the total power, respectively, and $G^{\mathrm{HV}}$ is their normalized correlation. The degree of polarization is $D_{\mathrm{p}}=|\lambda_{\mathrm{H}}-\lambda_{\mathrm{V}}|$, where $\lambda_{\mathrm{H}}$ and $\lambda_{\mathrm{V}}$ are the eigenvalues of $\mathbf{G}_{\mathrm{p}}$ \cite{Al-Qasimi07OL}. Spatial coherence can be similarly described via a spatial coherency matrix for points $\vec{a}$ and $\vec{b}$, $\mathbf{G}_{\mathrm{s}}=\left(\small{\begin{array}{cc}G_{aa}&G_{ab}\\G_{ba}&G_{bb}\end{array}}\right)$, where $G_{kl}=\langle E_{k}E_{l}^{*}\rangle$, $k,l=a,b$. The properties of $\mathbf{G}_{\mathrm{s}}$ are similar to those of $\mathbf{G}_{\mathrm{p}}$. The visibility of the interference fringes observed by superposing the fields from $\vec{a}$ and $\vec{b}$ is $V=2|G_{ab}|$. Alternatively, the degree of spatial coherence $D_{\mathrm{s}}=|\lambda_{a}-\lambda_{b}|$ represents the \textit{maximum} visibility obtained after equalizing the amplitudes at $\vec{a}$ and $\vec{b}$, where $\lambda_{a}$ and $\lambda_{b}$ are the eigenvalues of $\mathbf{G}_{\mathrm{s}}$ \cite{Zernike38P,Abouraddy17OE}.

A DoF represented by a $2\times2$ coherency matrix carries up to 1~bit of entropy; e.g., the polarization entropy is $S_{\mathrm{p}}=-\lambda_{\mathrm{H}}\log_{2}\lambda_{\mathrm{H}}-\lambda_{\mathrm{V}}\log_{2}\lambda_{\mathrm{V}}$, where $0\leq S_{\mathrm{p}}\leq1$. The zero-entropy state $S_{\mathrm{p}}=0$ corresponds to a fully polarized field (no statistical fluctuations), whereas the maximal-entropy state $S_{\mathrm{p}}=1$ corresponds to an unpolarized field (maximal fluctuations) \cite{Brosseau06PO}; similarly for the spatial DoF based on $\mathbf{G}_{\mathrm{s}}$. Entropy so defined is a \textit{unitary invariant} of the field DoF: it cannot be changed by applying lossless deterministic optical transformations.

\subsection{Joint polarization and spatial coherence formalism}

Evaluating $\mathbf{G}_{\mathrm{p}}$ and $\mathbf{G}_{\mathrm{s}}$ is \textit{not} sufficient to completely identify the coherence of a vector field in which the polarization and spatial DoFs are potentially correlated. A $4\times4$ coherency matrix $\mathbf{G}$ is necessary to capture the full vector-field coherence \cite{Gori06OL,Kagalwala13NP},
\begin{equation}
\mathbf{G}=\left(\begin{array}{cccc}
G_{aa}^{\mathrm{HH}}&G_{aa}^{\mathrm{HV}}&G_{ab}^{\mathrm{HH}}&G_{ab}^{\mathrm{HV}}\\[.4em]
G_{aa}^{\mathrm{VH}}&G_{aa}^{\mathrm{VV}}&G_{ab}^{\mathrm{VH}}&G_{ab}^{\mathrm{VV}}\\[.4em]
G_{ba}^{\mathrm{HH}}&G_{ba}^{\mathrm{HV}}&G_{bb}^{\mathrm{HH}}&G_{bb}^{\mathrm{HV}}\\[.4em]
G_{ba}^{\mathrm{VH}}&G_{ba}^{\mathrm{VV}}&G_{bb}^{\mathrm{VH}}&G_{bb}^{\mathrm{VV}}
\end{array}\right), \nonumber
\end{equation}
where $G_{kl}^{ij}=\langle E_{k}^{i}(E_{l}^{j})^{*}\rangle$, $i,j=\mathrm{H},\mathrm{V}$, and $k,l=a,b$. The matrix $\mathbf{G}$ is Hermitian positive semi-definite and normalized such that $\mathrm{Tr}\{\mathbf{G}\}=1$ (`$\mathrm{Tr}$' is the matrix trace). The diagonal elements are the power-fractions from the mutually exclusive contributions: $G_{aa}^{\mathrm{HH}}$ and $G_{aa}^{\mathrm{VV}}$ are the H and V components at $\vec{a}$, respectively, and $G_{bb}^{\mathrm{HH}}$ and $G_{bb}^{\mathrm{VV}}$ are those at $\vec{b}$. The off-diagonal elements are normalized correlations between field components. The double-slit visibility observed when overlapping the fields from $\vec{a}$ and $\vec{b}$ is $V=2|G_{ab}^{\mathrm{HH}}+G_{ab}^{\mathrm{VV}}|$ \cite{Wolf03PLA}. Crucially, $V$ is \textit{not a unitary invariant} of the field \cite{Tervo03OE,Setala04OE}, and reversible optical transformations that span the spatial and polarization DoFs can increase $V$ \cite{Gori07OL,Herrero07OL}.

Each \textit{physically independent} DoF (spatial and polarization) carries one bit of entropy, so the vector field now carries 2 bits of entropy: $S=-\lambda_{a\mathrm{H}}\log_{2}\lambda_{a\mathrm{H}}-\lambda_{a\mathrm{V}}\log_{2}\lambda_{a\mathrm{V}}-\lambda_{b\mathrm{H}}\log_{2}\lambda_{b\mathrm{H}}-\lambda_{b\mathrm{V}}\log_{2}\lambda_{b\mathrm{V}}$, where $0\leq S\leq2$ and $\{\lambda\}=\{\lambda_{a\mathrm{H}},\lambda_{a\mathrm{V}},\lambda_{b\mathrm{H}},\lambda_{b\mathrm{V}}\}$ are the real positive eigenvalues of $\mathbf{G}$. The zero-entropy state $S=0$ corresponds to a fully polarized \textit{and} spatially coherent field (no statistical fluctuations in either DoF and $\{\lambda\}=\{1,0,0,0\}$), whereas the maximal-entropy state $S=2$ corresponds to an unpolarized spatially incoherent field (maximal fluctuations in both DoFs and $\{\lambda\}=\tfrac{1}{4}\{1,1,1,1\}$).

\section{Concept of optical coherency conversion}

In general $S\leq S_{\mathrm{p}}+S_{\mathrm{s}}$, with equality achieved only when the two DoFs are independent, in which case $\mathbf{G}$ can be written in separable form $\mathbf{G}=\mathbf{G}_{\mathrm{s}}\otimes\mathbf{G}_{\mathrm{p}}$. In general, $S_{\mathrm{s}}$ and $S_{\mathrm{p}}$ are obtained from the $2\times2$ `reduced' spatial and polarization coherency matrices
\begin{align}
\mathbf{G}_{\mathrm{s}}^{(\mathrm{r})}=\left(\begin{array}{cc}
G_{aa}^{\mathrm{HH}}+G_{aa}^{\mathrm{VV}}
&G_{ab}^{\mathrm{HH}}+G_{ab}^{\mathrm{VV}}\\ [0.4em]
G_{ba}^{\mathrm{HH}}+G_{ba}^{\mathrm{VV}}
&G_{bb}^{\mathrm{HH}}+G_{bb}^{\mathrm{VV}}
\end{array}\right), \nonumber  \\ 
 \mathbf{G}_{\mathrm{p}}^{(\mathrm{r})}=\left(\begin{array}{cc}
G_{aa}^{\mathrm{HH}}+G_{bb}^{\mathrm{HH}}
&G_{aa}^{\mathrm{HV}}+G_{bb}^{\mathrm{HV}}\\ [0.4em]
G_{aa}^{\mathrm{VH}}+G_{aa}^{\mathrm{VH}}
&G_{aa}^{\mathrm{VV}}+G_{bb}^{\mathrm{VV}}
\end{array}\right), \nonumber  
\end{align}
which are obtained from $\mathbf{G}$ by a `partial trace' \cite{Peres95Book}, that is, by tracing over one DoF \cite{Kagalwala13NP,Abouraddy14OL}

\begin{figure}[hb!]
\centering\includegraphics[scale=1]{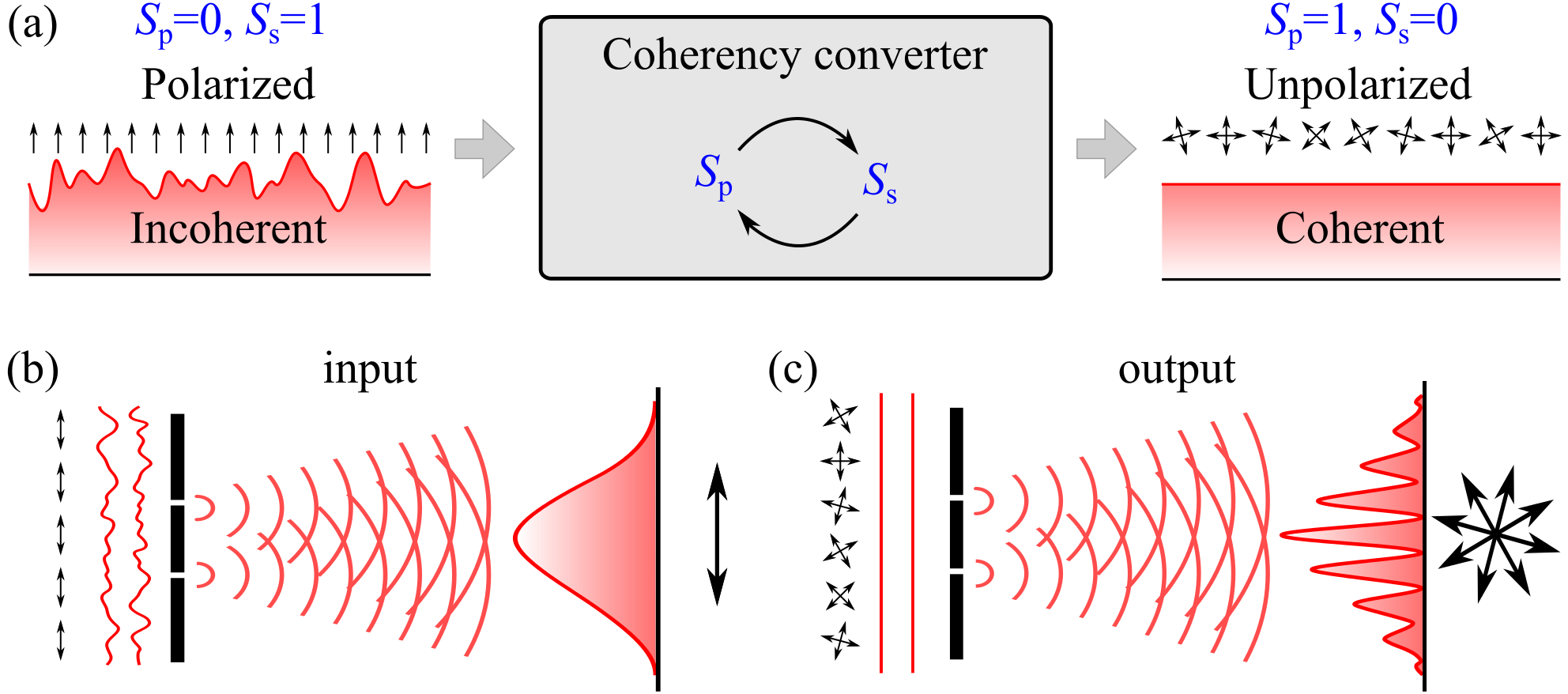}
\caption{Concept of an optical-coherence converter. 
(a) Starting with a polarized but spatially incoherent field ($S_{\mathrm{p}}=0$ and $S_{\mathrm{s}}=1$, $S=S_{\mathrm{p}}+S_{\mathrm{s}}=1$), coherence is converted from polarization to the spatial DoF, thereby yielding an unpolarized but spatially coherent field ($S_{\mathrm{p}}=1$ and $S_{\mathrm{s}}=0$) but without introducing further fluctuations (fixed total entropy $S=1$). The device thus converts the statistical fluctuations (and the attendant entropy) from one DoF to the other. (b) When a polarized but spatially incoherent field is incident on a double-slit, no interference fringes are observed. (c) After converting coherence from polarization to the spatial DoF, high-visibility (but unpolarized) interference fringes appear.}\label{fig:GeneralConcept}
\end{figure}

The concept of an optical-coherence converter is illustrated in Fig.~\ref{fig:GeneralConcept}(a). Consider the case when the field carries \textit{one bit} of entropy ($S=1$) and the DoFs are independent ($S=S_{\mathrm{p}}+S_{\mathrm{s}}$), in which case a single DoF can accommodate this entropy. The field may be maximally \textit{incoherent} but polarized ($S_{\mathrm{s}}=1$ and $S_{\mathrm{p}}=0$), whereupon no interference fringes can be observed [Fig.~\ref{fig:GeneralConcept}(b)]. Alternatively, the field may be spatially coherent but \textit{un}polarized ($S_{\mathrm{s}}=0$ and $S_{\mathrm{p}}=1$), in which case full-visibility fringes can be observed [Fig.~\ref{fig:GeneralConcept}(c)]. We demonstrate here that an optical field can be \textit{reversibly} transformed from the former configuration to the latter without loss of energy, thus \textit{converting} coherence from one DoF (polarization) to the other (spatial). Throughout the procedure $S$ remains constant; that is, no uncertainty is added or removed from the field, only an internal reorganization of the field engendered by a unitary transformation confines the statistical fluctuations to one DoF while freeing the other from uncertainty. We call such a system a `coherence converter'.

The optical arrangement we propose to convert coherency between the spatial and polarization DoFs is depicted in Fig.~\ref{fig:SetupExp1and2}. We start from two points $\vec{a}'$ and $\vec{b}'$ of equal intensity in a spatially \textit{in}coherent H-polarized field (the fields are mutually incoherent or statistically independent), which thus produce no interference fringes. The polarization at $\vec{b}'$ is rotated to become orthogonal to that at $\vec{a}'$ $(\mathrm{H}\rightarrow\mathrm{V})$ before combining the fields at a polarizing beam splitter (PBS), which yields an \textit{un}polarized field. We then split the field into two points $\vec{a}$ and $\vec{b}$ using a \textit{non}polarizing beam splitter, which creates two copies of the field that can demonstrate high-visibility interference fringes. We proceed now to present the measurements at each step of this coherency-conversion process.

\section{Source characterization}

\begin{figure}[t!]
\centering\includegraphics[scale=1]{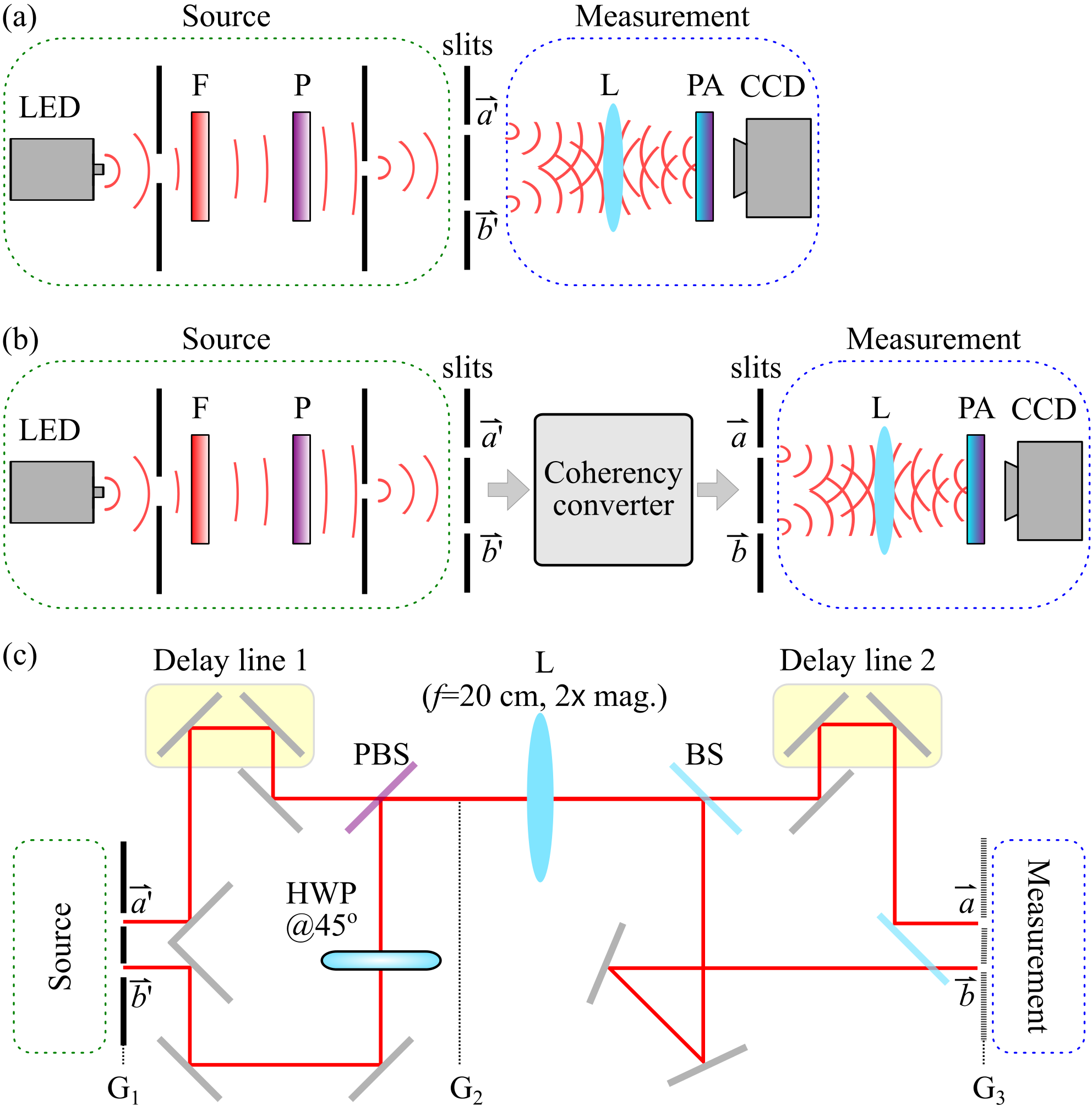}
\caption{(a) Schematic depicting the input field preparation (source) and characterization (measurement). The field at points $\vec{a}'$ and $\vec{b}'$ is spatially incoherent but fully polarized (scalar). F: Filter; P: polarizer; L: lens; PA: polarization analyzer; CCD: charge-coupled device camera. (b) A coherency converter maps the spatially incoherent but polarized field at $\vec{a}'$ and $\vec{b}'$ to a spatially coherent but unpolarized field at $\vec{a}$ and $\vec{b}$. (c) Schematic of the optical setup for the coherence-converter. A bi-convex lens (L: $f=20$~cm)  images $\vec{a}'$ and $\vec{b}'$ to $\vec{a}$ and $\vec{b}$, respectively, with $2\times$ magnification. The delay lines enable matching pairs of paths within the source temporal coherence length. HWP: Half-wave plate; PBS: polarizing beam splitter; BS: beam splitter. The planes at which the coherency matrices $\mathbf{G}_{1}$, $\mathbf{G}_{2}$, and $\mathbf{G}_{3}$ are reconstructed are marked.}\label{fig:SetupExp1and2} 
\end{figure}

\begin{figure}[t!]
\centering\includegraphics[scale=1]{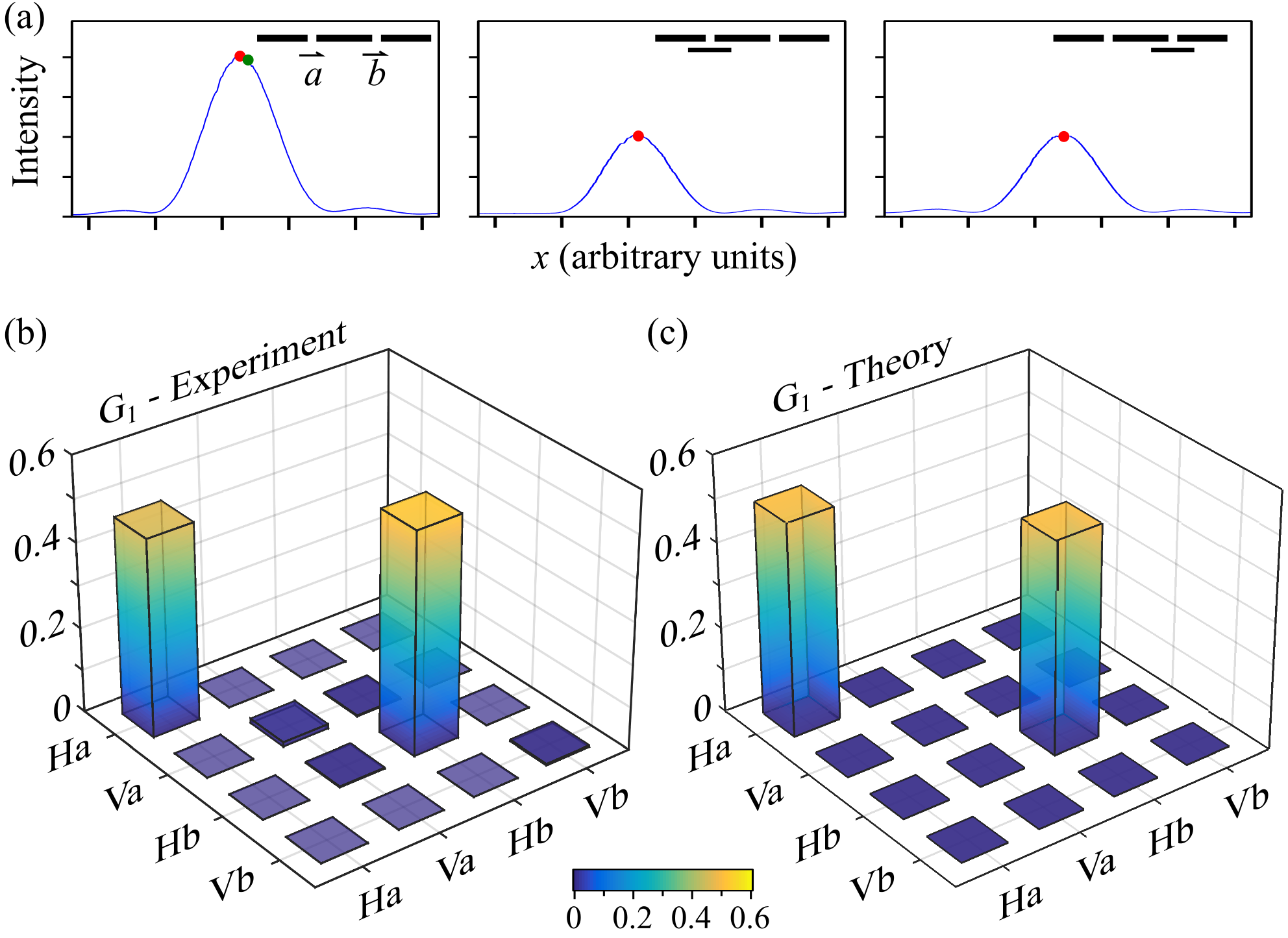}
\caption{(a) The four measurements required to reconstruct the spatial coherence matrix $\mathbf{G}_{\mathrm{S}}$ for a scalar field at $\vec{a}$ and $\vec{b}$. The intensity pattern is recorded with both slits open (left), and two measurements are made: the intensity on the optical axis (red dot) and at the location mid-way along the first expected fringe location calculated from the slit separation (green dot). No fringes are observed here since the field is spatially incoherent. Next, the intensity on the optical axis is recorded when $\vec{a}$ (left) and $\vec{b}$ (right) are blocked (the red dots; see Refs.~\cite{Abouraddy14OL,Kagalwala15SR} for details). (b) Plot depicting graphically the real parts of the elements of the spatial-polarization coherency matrix $\mathbf{G}_{1}$ for the source plane as reconstructed from OCmT that utilizes the measurements in (a) when carried out in conjunction with polarization measurements. (c) Plot depicting graphically the elements of the theoretically expected coherency matrix $\mathbf{G}=\tfrac{1}{2}\mathrm{diag}\{1,0,1,0\}$, corresponding to a scalar H-polarized field that is spatially incoherent (Eq.~\ref{G_1}).}\label{fig:Figure3}
\end{figure}

The optical field we study is extracted from a broadband, unpolarized LED (center wavelength 850~nm, 30-nm-FWHM bandwidth; Thorlabs M850L3 IR). The field is spectrally filtered (10-nm-FWHM), polarized along H, and spatially filtered through a 100-$\mu$m-wide slit placed at a distance of 180~mm from the source. The `input' plane that includes the points $\vec{a}'$ and $\vec{b}'$ (each defined by a 100-$\mu$m-wide slit) is located 420~mm away from the slit [the source in Fig.~\ref{fig:SetupExp1and2}(a)]. We first confirm that the field is spatially coherent within $\vec{a}'$ and $\vec{b}'$ separately (i.e., the spatial coherence width of the field, estimated to be $\sim$1 mm, is larger than the slit width). This is accomplished using a narrow pair of slits (50-$\mu$m-wide separated by $\Delta=150$~$\mu$m) at either $\vec{a}'$ or $\vec{b}'$ and observing the double-slit interference on a CCD camera (Hamamatsu 1394) at a distance of $d=200$~mm away. High-visibility fringes ($V=0.98$) are observed separated by $\lambda d/\Delta\approx1170$~$\mu$m. Next, we superpose the fields from $\vec{a}'$ and $\vec{b}'$ ['measurement' in Fig.~\ref{fig:SetupExp1and2}(a)] and observe no interference fringes [Fig.~\ref{fig:Figure3}(a)], confirming that the two points are separated by more than the field coherence width. We have thus confirmed the relationship between the two length scales involved: the size of the locations at $\vec{a}'$ and $\vec{b}'$ (100~$\mu$m) is smaller than the coherence width, and the separation between them (10~mm) is larger. The field that we start from is linearly polarized (scalar) but spatially incoherent, thus $\mathbf{G}$ has the form
\begin{align}\label{G_1}
\mathbf{G}_{1}&=\tfrac{1}{2}\left(\begin{array}{cccc}
1&0&0&0\\
0&0&0&0\\
0&0&1&0\\
0&0&0&0
\end{array}\right)=\tfrac{1}{2}\mathrm{diag}\{1,0,1,0\} \nonumber \\
&=\tfrac{1}{2}\left(\begin{array}{cc}1&0\\0&1\end{array}\right)_{\mathrm{s}}\otimes\left(\begin{array}{cc}1&0\\0&0\end{array}\right)_{\mathrm{p}},
\end{align}
where the subscripts `s' and `p' refer to the spatial and polarization DoFs, respectively, and the notation $\mathrm{diag}\{\dot\}$ identifies a diagonal matrix with the entries along the diagonal listed between the curly brackets.

To fully characterize the field coherence across the spatial and polarization DoFs, we measure $\mathbf{G}_{1}$ via OCmT \cite{Abouraddy14OL,Kagalwala15SR}, which requires 16 measurements to reconstruct $\mathbf{G}_{1}$. Since 4 polarization projections are required to identify $\mathbf{G}_{\mathrm{p}}$ and 4 spatial projections are required to determine $\mathbf{G}_{\mathrm{s}}$ for a scalar field, $4\times4$ linearly independent combinations of these spatial and polarization projections are necessary to reconstruct $\mathbf{G}$ subject to the constraints of Hermiticity, semi-positiveness, and unity-trace. These measurements are in one-to-one correspondence to those required to reconstruct a two-qubit density matrix in quantum mechanics, a process known as `quantum state tomography' \cite{Wooters90CEPI,James01PRA1,Abouraddy02OptComm}. Carrying out these optical measurements (see \cite{Kagalwala15SR} for details), $\mathbf{G}_{1}$ is reconstructed [Fig.~\ref{fig:Figure3}(b)] and is found to be in good agreement with the theoretical expectation [Fig.~\ref{fig:Figure3}(c)], with the remaining slight deviations attributable to unequal powers at $\vec{a}'$ and $\vec{b}'$.

The measured coherency matrix $\mathbf{G}_{1}$ in the $(\vec{a}',\vec{b}')$-plane yields $S=1.001$, and the reduced spatial and polarization coherency matrices $\mathbf{G}_{\mathrm{s}}^{(\mathrm{r})}$ and $\mathbf{G}_{\mathrm{p}}^{(\mathrm{r})}$ obtained from $\mathbf{G}$ yield $S_{\mathrm{s}}=0.991$, $S_{\mathrm{p}}=0.037$, respectively. The field entropy is thus associated with the spatial DoF and \textit{not} polarization, resulting in an absence of interference fringes [Fig.~\ref{fig:Figure3}(a)]. The lack of interference fringes is consistent with the fact that all measures of spatial coherence or double-slit interference fringes for a \textit{vector field} rely on the cross-correlation matrix \cite{Wolf03PLA} or beam-coherence matrix \cite{Gori98OL}, $\mathbf{G}_{ab}=\footnotesize{\left(\begin{array}{cc}G_{ab}^{\mathrm{HH}}&G_{ab}^{\mathrm{HV}}\\G_{ab}^{\mathrm{VH}}&G_{ab}^{\mathrm{VV}}\end{array}\right)}$, which is the top right $2\times2$ block of the coherency matrix $\mathbf{G}$. For example, the degree of coherence proposed by E. Wolf \cite{Wolf03PLA}, the degree of electromagnetic coherence \cite{Tervo03OE,Setala04OE}, the complex degree of mutual polarization \cite{Ellis04OL}, the visibility predicted through a generalized form of the Fresnel-Arago law \cite{Mujat04JOSAA}, or the maximal visibility obtainable through local unitary transformations \cite{Gori07OL,Herrero07OL} all predict that the observed visibility will be zero if all the entries in $\mathbf{G}_{ab}$ are zero. Because the measured and theoretically expected $\mathbf{G}_{1}$ has all zero entries in the $\mathbf{G}_{ab}$ block, it follows that interference fringes do not form. We proceed to show that interference fringes nevertheless appear once the coherence has been \textit{reversibly converted} from polarization to the spatial DoF.

\begin{figure}[t!]
\centering\includegraphics[scale=1]{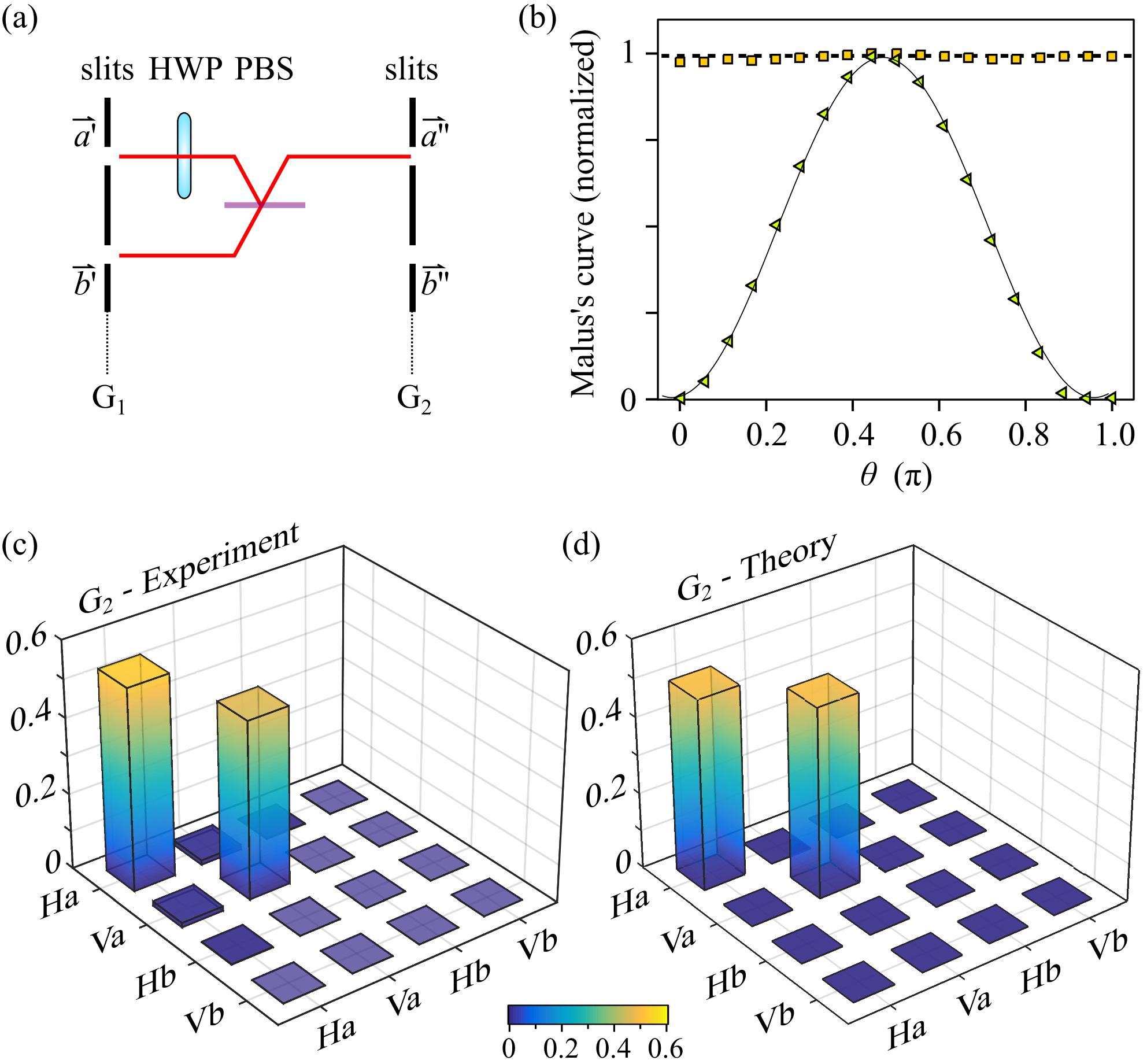}
\caption{(a) Schematic for the setup to combine two linearly polarized fields from $\vec{a}'$ and $\vec{b}'$ that are statistically independent or spatially incoherent ($S_{\mathrm{p}}=0$ and $S_{\mathrm{s}}=1$) into $\vec{a}''$ and $\vec{b}''$ whereupon the field becomes unpolarized but spatially coherent ($S_{\mathrm{p}}=1$ and $S_{\mathrm{s}}=0$), without loss of power or increase in total entropy $S=1$. HWP: Half-wave plate rotated to implement the transformation H$\rightarrow$V; PBS: polarizing beam splitter. (b) Malus curves for fields at the two input ports of the PBS highlight the linear polarization (one orthogonal to the other) and that for the field at the output port highlights its random polarization. The dashed and continuous lines are the flat and sinusoidal curves associated with unpolarized and V-polarized light, respectively. (c) Graphical depiction of the elements of the full coherency matrix $\mathbf{G}_{2}$ is obtained experimentally and (d) expected theoretically.}\label{Figure4}
\end{figure}

\begin{figure*}[t!]
\centering\includegraphics[scale=1]{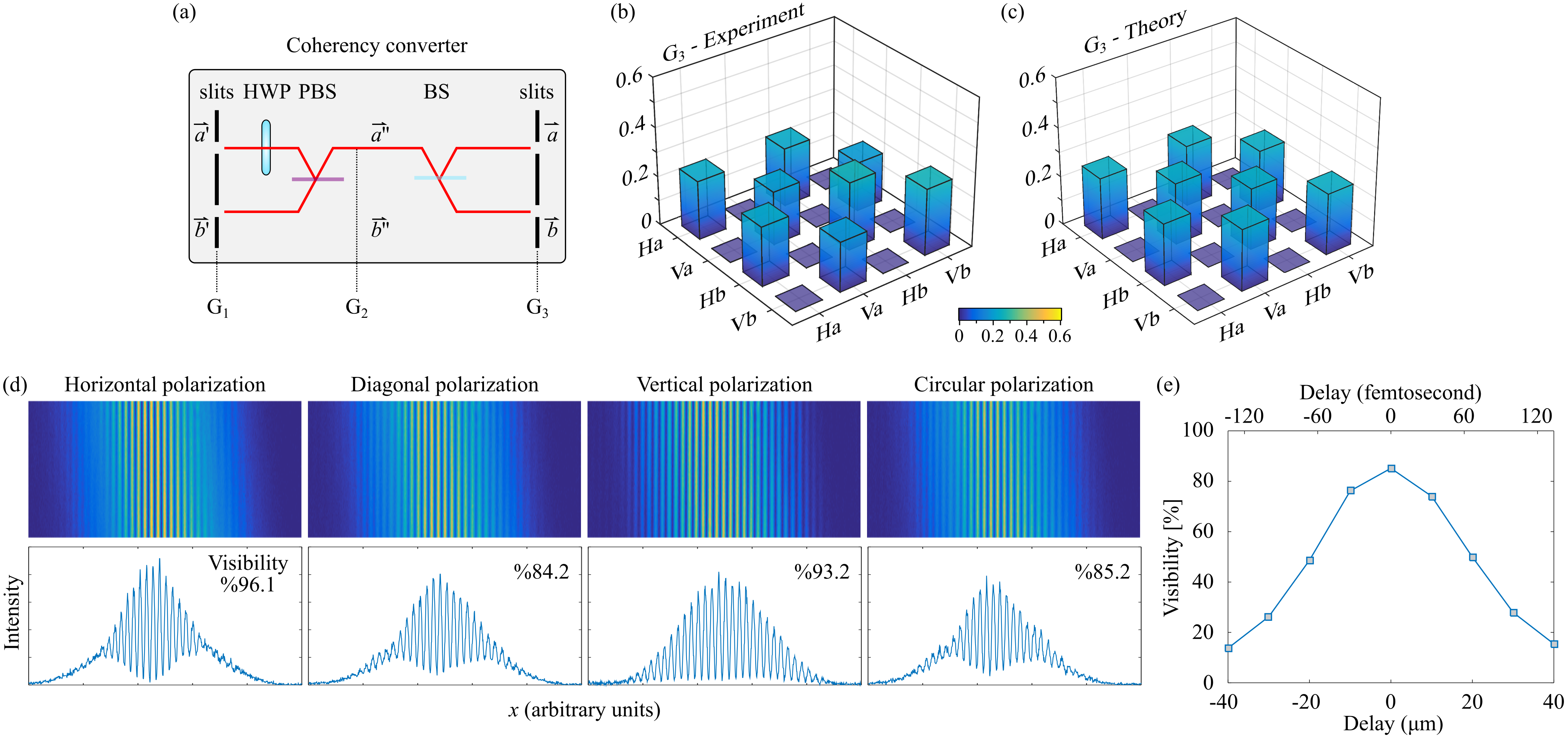}
\caption{(a) Schematic for the coherence converter that transforms two linearly polarized, spatially incoherent fields (at $\vec{a}'$ and $\vec{b}'$) into two randomly polarized mutually coherent fields (at $\vec{a}$ and $\vec{b}$). (b) Graphical depiction of the real part of the entries of the experimentally reconstructed $\mathbf{G}_{3}$ via OCmT. (c) The theoretical expectation for $\mathbf{G}_{3}$. (d) Interference patterns obtained by overlapping the fields from $\vec{a}$ and $\vec{b}$ after a polarization projection, with high-visibility fringes observed in all cases. The top panels are CCD camera images and the lower panels are obtained by integrating the fringes vertically.(e) Visibility as a function of a relative delay inserted between the fields at $\vec{a}$ and $\vec{b}$ before overlapping them at the CCD camera [Fig.~\ref{fig:SetupExp1and2}(c)] for the diagonal polarization projection case in (d).}\label{Figure5}
\end{figure*}

\section{Converting coherence from polarization to space}

Converting the coherence reversibly from polarization to the spatial DoF entails maximizing the polarization entropy $S_{\mathrm{p}}$ and minimizing the spatial entropy $S_{\mathrm{s}}$ at \textit{fixed} total entropy $S$. A half-wave plate (HWP) placed after $\vec{b}'$ rotates the polarization H$\rightarrow$V, resulting in a new coherency matrix $\mathbf{G}=\tfrac{1}{2}\mathrm{diag}\{1,0,0,1\}$. The fields from $\vec{a}'$ and $\vec{b}'$ are then directed by mirrors to the two input ports of a PBS (Thorlabs CM1-PBS252), where the V component is transmitted and H is reflected [Fig.~\ref{Figure4}(a)]. Consequently, the H and V components overlap in the same output port at $\vec{a}''$ (minimal power in the other port $\vec{b}''$). Note that a PBS is a reversible device: when the two output fields reverse their direction and return to the PBS, the input fields are reconstituted. The field is now unpolarized, which we confirm by registering a flat Malus curve and comparing it to the sinusoidal Malus curve produced by the linearly polarized input fields [Fig.~\ref{Figure4}(b)]. That is, each incident field on the PBS is linearly polarized, whereas their superposition at the output -- with the initial optical power now concentrated in a single path -- is randomly polarized. The coherency matrix $\mathbf{G}_{2}$ at $\vec{a}''$ and $\vec{b}''$ is reconstructed via OCmT [Fig.~\ref{Figure4}(c)], and is found to be in good agreement with the expected form [Fig.~\ref{Figure4}(d)]:
\begin{align}
\mathbf{G}_{2}&=\tfrac{1}{2}\left(\begin{array}{cccc}
1&0&0&0\\
0&1&0&0\\
0&0&0&0\\
0&0&0&0
\end{array}\right)=\tfrac{1}{2}\mathrm{diag}\{1,1,0,0\} \nonumber \\ 
&=\left(\begin{array}{cc}1&0\\0&0\end{array}\right)_{\mathrm{s}}\otimes\tfrac{1}{2}\left(\begin{array}{cc}1&0\\0&1\end{array}\right)_{\mathrm{p}}.
\end{align}

From the reconstructed $\mathbf{G}_{2}$ in the $(\vec{a}'',\vec{b}'')$-plane, the total entropy is $S=0.997$, but the new values of entropy for the spatial and polarization DoFs are $S_{\mathrm{s}}=0.001$, and $S_{\mathrm{p}}=0.996$, respectively. \textit{The entropy has now been converted between the two DoFs}. To observe interference fringes, the randomly polarized field at $\vec{a}''$ is split symmetrically into two halves by a 50:50 \textit{non}-polarizing beam splitter to points $\vec{a}$ and $\vec{b}$ [Fig.~\ref{Figure5}(a)], which can then be overlapped to produce high-visibility fringes. This step does not change the values of $S$, $S_{\mathrm{s}}$, or $S_{\mathrm{p}}$. The $(\vec{a},\vec{b})$-plane is the image plane relayed from the $(\vec{a}',\vec{b}')$-plane by a lens [Fig.~\ref{fig:SetupExp1and2}(c)]. The coherency matrix $\mathbf{G}_{3}$ in this plane: 
\begin{equation}\label{G_3}
\mathbf{G}_{3}=\frac{1}{4}\left(\begin{array}{cccc}
1&0&1&0\\
0&1&0&1\\
1&0&1&0\\
0&1&0&1
\end{array}\right)=\frac{1}{2}\left(\begin{array}{cc}1&1\\1&1\end{array}\right)_{\mathrm{s}}\otimes\frac{1}{2}\left(\begin{array}{cc}1&0\\0&1\end{array}\right)_{\mathrm{p}}.
\end{equation}
The measured $\mathbf{G}_{3}$ reconstructed via OCmT [Fig.~\ref{Figure5}(b)] is in good agreement with the theoretical expectation [Fig.~\ref{Figure5}(c)].

Now that we have $G_{ab}^{\mathrm{HH}}+G_{ab}^{\mathrm{VV}}=\tfrac{1}{2}$, the predicted visibility is $V=1$ \cite{Wolf03PLA}. Furthermore, because the two DoFs are independent \textit{and} the field is unpolarized (as is clear from the separable form of $\mathbf{G}_{3}$ in Eq.~\ref{G_3}), projecting the polarization on any direction will not affect the high visibility [Fig.~\ref{Figure5}(d)]. Note that the coherence time of the field is determined by its spectral bandwidth, and observing the fringes [Fig.~\ref{Figure5}(d)] requires introducing a relative delay in the path of the fields from $\vec{a}$ or $\vec{b}$ before overlapping them [delay line 2 in Fig.~\ref{fig:SetupExp1and2}(c)]. The variation in the measured visibility with introduced relative time delay corresponds to the expected field coherence time $\approx100$~fs (coherence length $\approx30$~$\mu$m) [Fig.~\ref{Figure5}(e)].

\section{Surviving randomization through entropy reallocation}

The ability to redistribute or reallocate the field entropy $S$ between the DoFs can be exploited in protecting a DoF from the deleterious impact of a randomizing medium. Consider a depolarizing medium represented by a Mueller matrix $\hat{M}=\mathrm{diag}\{1,0,0,0\}$ that converts \textit{any} state of polarization into a completely unpolarized state. The initial field $\mathbf{G}_{1}$ [Fig.~\ref{fig:Figure3}(b,c)] would be converted into the incoherent unpolarized field $\mathbf{G}'_{1}=\tfrac{1}{4}\mathrm{diag}\{1,1,1,1\}$ upon traversing this medium with $S_{\mathrm{p}}\rightarrow1$, such that $S\rightarrow2$. If, however, coherence is first reversibly converted from the spatial DoF to polarization ($S_{\mathrm{p}}\rightarrow1$ and $S_{\mathrm{s}}\rightarrow0$), then traversing a depolarizing medium \textit{cannot increase} $S_{\mathrm{p}}$, and the field is thus left unchanged. Subsequently, the coherence-conversion can be reversed and a polarized field retrieved after emerging from the depolarizing medium without loss of energy.

\begin{figure}[t!]
\centering\includegraphics[scale=1]{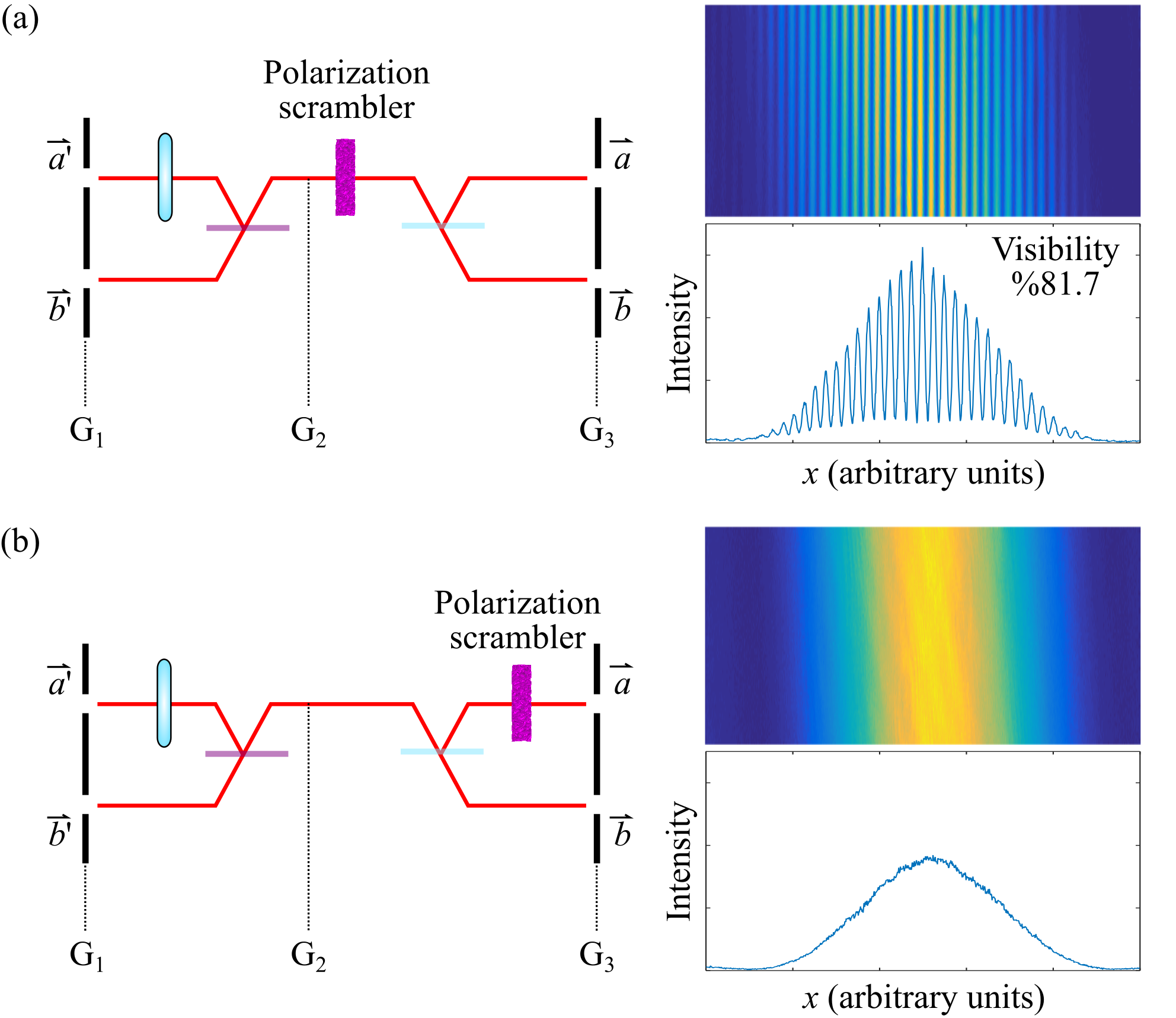}
\caption{The effect of a polarization scrambler on the field when introduced at two different planes. (a) A polarization scrambler is placed at the $\mathbf{G}_{2}$ plane (after the PBS at $\vec{a}''$) has no effect on the visibility of the interference pattern measured at the $\mathbf{G}_{3}$ plane, as shown in the right panels. (b) A polarization scrambler placed before $\vec{a}$ destroys the visibility. (a,b) The measurements on the right panels are averaged over the polarization shown in Fig.~\ref{Figure5}(d).}\label{Figure6}
\end{figure}

We have carried out the proof-of-concept experiment depicted in Fig.~\ref{Figure6}(a) where we place a depolarizer or polarization scrambler in the path of the field $\mathbf{G}_{2}$ in the $(\vec{a}'',\vec{b}'')$-plane. The polarization scrambler is implemented by rotating a HWP. The CCD camera exposure time is increased to 10~s, corresponding to the rotation time of the waveplate, to capture the averaged interference pattern in a single shot. The Mueller matrix associated with a polarization scrambler is $\hat{M}=\mathrm{diag}\{1,0,0,0\}$. The visibility observed after the $(\vec{a},\vec{b})$-plane remains high. This result can be modeled theoretically by first noting that a unitary transformation $\hat{U}$ transforms the field according to $\mathbf{G}_{2}\rightarrow\mathbf{G}'_{2}=\hat{U}\mathbf{G}_{2}\hat{U}^{\dag}$, where $\hat{U}$ is a $4\times4$ unitary transformation that spans the spatial and polarization DoFs. If the device in question impacts the two DoFs independently, then $\hat{U}=\hat{U}_{\mathrm{s}}\otimes\hat{U}_{\mathrm{p}}$, where $\hat{U}_{\mathrm{s}}$ and $\hat{U}_{\mathrm{p}}$ are $2\times2$ unitary transformations for the spatial and polarization DoFs, respectively. The impact of a randomizing but energy-conserving transformation can be modeled as a statistical ensemble of unitary transformations \cite{Kim87JOSAA,Gil00JOSAA,Gamel11OL}. The transformation of $\mathbf{G}_{2}$ upon traversing a polarization scrambler can be expressed as
\begin{equation}\label{G2AfterScrambler}
\mathbf{G}_{2}'=\sum_{j=1}^{4}p_{j}\{\hat{\mathbb{I}}\otimes\hat{U}_{\mathrm{p}}^{(j)}\}\mathbf{G}_{2}\{\hat{\mathbb{I}}\otimes\hat{U}_{\mathrm{p}}^{(j)\dag}\},
\end{equation}
where $\hat{\mathbb{I}}$ is the $2\times2$ identity matrix and the ensemble $\{U_{\mathrm{p}}^{(j)}\}$ comprises with equal probabilities $p_{j}=\tfrac{1}{4}$ the Jones matrices: $\tfrac{1}{\sqrt{2}}\footnotesize{\left(\begin{array}{cc}1&-1\\1&1\end{array}\right)}$,
$\tfrac{1}{\sqrt{2}}\footnotesize{\left(\begin{array}{cc}1&1\\-1&1\end{array}\right)}$,
$\footnotesize{\left(\begin{array}{cc}1&0\\0&-1\end{array}\right)}$,
$\footnotesize{\left(\begin{array}{cc}0&1\\1&0\end{array}\right)}$, which correspond to polarization rotations of $\pm45^{\circ}$ in the H-V basis, and a HWP in the H-V basis and rotated by $45^{\circ}$. Substituting the ensemble $\{\hat{U}_{\mathrm{p}}^{(j)}\}$ into Eq.~\ref{G2AfterScrambler} yields $\mathbf{G}_{2}'=\mathbf{G}_{2}$, which entails that the high-visibility seen in Fig.~\ref{Figure5}(d) should be retained, as confirmed in Fig.~\ref{Figure6}(a). In other words, $\mathbf{G}_{2}$ is \textit{invariant with regards to any polarization randomization}.

If the polarization scrambler is position-dependent, the coherency matrix $\mathbf{G}_{2}$ will no longer be invariant under randomization (because the spatial and polarization DoFs become coupled). In the experiment illustrated in Fig.~\ref{Figure6}(b), the polarization scrambler is placed at $\vec{a}$ in the plane of $\mathbf{G}_{3}$. The spatial-polarization transformation of $\mathbf{G}_{3}$ takes the form
\begin{align}\label{G3AfterScrambler}
\mathbf{G}_{3}'=&\{\hat{\Lambda}_{b}\otimes\mathbb{I}\}\mathbf{G}_{3}\{\hat{\Lambda}_{b}^{\dag}\otimes\mathbb{I}\} \nonumber  \\
&+\sum_{j=1}^{4}p_{j}
\left\{\hat{\Lambda}_{a}\otimes\hat{U}_{\mathrm{p}}^{(j)}\right\}\mathbf{G}_{3}\left\{\hat{\Lambda}_{a}^{\dag}\otimes\hat{U}_{\mathrm{p}}^{(j)\dag}\right\},
\end{align}
where we have $\hat{\Lambda}_{a}=\footnotesize{\left(\begin{array}{cc}1&0\\0&0\end{array}\right)}$ and $\hat{\Lambda}_{b}=\footnotesize{\left(\begin{array}{cc}0&0\\0&1\end{array}\right)}$, resulting in $\mathbf{G}_{3}'=\tfrac{1}{2}\hat{\mathbb{I}}\otimes\tfrac{1}{2}\hat{\mathbb{I}}=\tfrac{1}{4}\mathrm{diag}\{1,1,1,1\}$; that is, an unpolarized and spatially incoherent field with maximal entropy $S=2$ is produced. No fringes will appear in this case [Fig.~\ref{Figure6}(b)].

\section{Discussion and conclusion}

To facilitate the analysis of the coherence of optical fields encompassing multiple DoFs, it is becoming increasingly clear that the mathematical formalism of multi-partite quantum mechanical states is most useful \cite{Simon10PRL,Qian11OL,Kagalwala13NP}. The underlying foundation for this utility is the analogy between the mathematical description used in these domains. The Hilbert space of a multi-partite system is the tensor product of the Hilbert spaces of the single-particle subsystems. In classical optics, the multiple DoFs of a beam are described in a linear vector space having formally the structure of a tensor product of the linear vector spaces of the individual DoFs. In the quantum setting, pure multi-partite states that cannot be separated into products of single-particle states are said to be entangled; whereas in the classical setting, fields in which the DoFs cannot be separated are now being called `classically entangled' \cite{Kagalwala13NP}, coherence can be viewed as a `resource' that may be reversibly converted from one DoF of the beam to another, just as entanglement is a resource shared among multiple quantum particles. There is a critical difference though between the quantum and classical settings. Entanglement between initially separable particles requires nonlocal operations of particle-particle interactions (which are particularly challenging for photons \cite{Gaeta13NP}); on the other hand, entangling operations that couple different DoFs of a beam are readily available in classical optics. Adopting this information-driven standpoint has led to a host of novel opportunities and applications. For example, Bell's measure, originally developed for ascertaining nonlocality, becomes a useful in quantifying the coherence of a multi-DoF beam and assessing the resources required to synthesize a beam of given characteristics \cite{Kagalwala13NP}; beams in which spatial modes and polarization are classically entangled have been shown to enable fast particle tracking \cite{Berg15Optica} and full Mueller characterization of a sample \cite{Toppel14NJP, Tripathi09OE}; and introducing spatio-temporal spectral correlations leads to propagation-invariant pulsed optical beams \cite{Kondakci16OE,Parker16OE}.

We have demonstrated -- for the first time to the best of our knowledge -- the `conversion' or transformation of coherence from one DoF of an optical field to another; namely, from polarization to the spatial DoF. Starting from a field that is spatially incoherent but polarized, we redirect the statistical fluctuations from space towards polarization, resulting in an unpolarized field that is spatially coherent -- reversibly, without losing optical energy along the way. Specifically, the 1 bit of entropy characterizing the spatial DoF was removed and added instead to the initially zero-entropy polarization. Entropy-engineering of partially coherent optical fields can open the path to a variety of possible future extensions and applications with regards to optimizing the interaction of optical fields with disordered media. \\

\noindent \textbf{Funding Information}. 
Office of Naval Research (ONR) contract N00014-14-1-0260. \\

\noindent  \textbf{Acknowledgments}.
The authors thank A.~Dogariu and D.~N.~Christodoulides for useful discussions.

\bibliography{sample}

\begin{thebibliography}{35}%
\makeatletter
\providecommand \@ifxundefined [1]{%
 \@ifx{#1\undefined}
}%
\providecommand \@ifnum [1]{%
 \ifnum #1\expandafter \@firstoftwo
 \else \expandafter \@secondoftwo
 \fi
}%
\providecommand \@ifx [1]{%
 \ifx #1\expandafter \@firstoftwo
 \else \expandafter \@secondoftwo
 \fi
}%
\providecommand \natexlab [1]{#1}%
\providecommand \enquote  [1]{``#1''}%
\providecommand \bibnamefont  [1]{#1}%
\providecommand \bibfnamefont [1]{#1}%
\providecommand \citenamefont [1]{#1}%
\providecommand \href@noop [0]{\@secondoftwo}%
\providecommand \href [0]{\begingroup \@sanitize@url \@href}%
\providecommand \@href[1]{\@@startlink{#1}\@@href}%
\providecommand \@@href[1]{\endgroup#1\@@endlink}%
\providecommand \@sanitize@url [0]{\catcode `\\12\catcode `\$12\catcode
  `\&12\catcode `\#12\catcode `\^12\catcode `\_12\catcode `\%12\relax}%
\providecommand \@@startlink[1]{}%
\providecommand \@@endlink[0]{}%
\providecommand \url  [0]{\begingroup\@sanitize@url \@url }%
\providecommand \@url [1]{\endgroup\@href {#1}{\urlprefix }}%
\providecommand \urlprefix  [0]{URL }%
\providecommand \Eprint [0]{\href }%
\providecommand \doibase [0]{http://dx.doi.org/}%
\providecommand \selectlanguage [0]{\@gobble}%
\providecommand \bibinfo  [0]{\@secondoftwo}%
\providecommand \bibfield  [0]{\@secondoftwo}%
\providecommand \translation [1]{[#1]}%
\providecommand \BibitemOpen [0]{}%
\providecommand \bibitemStop [0]{}%
\providecommand \bibitemNoStop [0]{.\EOS\space}%
\providecommand \EOS [0]{\spacefactor3000\relax}%
\providecommand \BibitemShut  [1]{\csname bibitem#1\endcsname}%
\let\auto@bib@innerbib\@empty
\bibitem [{\citenamefont {Mandel}\ and\ \citenamefont
  {Wolf}(1995)}]{Mandel95Book}%
  \BibitemOpen
  \bibfield  {author} {\bibinfo {author} {\bibfnamefont {L.}~\bibnamefont
  {Mandel}}\ and\ \bibinfo {author} {\bibfnamefont {E.}~\bibnamefont {Wolf}},\
  }\href@noop {} {\emph {\bibinfo {title} {Optical Coherence and Quantum
  Optics}}}\ (\bibinfo  {publisher} {Cambridge Univ. Press},\ \bibinfo
  {address} {Cambridge},\ \bibinfo {year} {1995})\BibitemShut {NoStop}%
\bibitem [{\citenamefont {Zernike}(1938)}]{Zernike38P}%
  \BibitemOpen
  \bibfield  {author} {\bibinfo {author} {\bibfnamefont {F.}~\bibnamefont
  {Zernike}},\ }\href@noop {} {\bibfield  {journal} {\bibinfo  {journal}
  {Physica}\ }\textbf {\bibinfo {volume} {5}},\ \bibinfo {pages} {785}
  (\bibinfo {year} {1938})}\BibitemShut {NoStop}%
\bibitem [{\citenamefont {Born}\ and\ \citenamefont {Wolf}(1999)}]{Born99Book}%
  \BibitemOpen
  \bibfield  {author} {\bibinfo {author} {\bibfnamefont {M.}~\bibnamefont
  {Born}}\ and\ \bibinfo {author} {\bibfnamefont {E.}~\bibnamefont {Wolf}},\
  }\href@noop {} {\emph {\bibinfo {title} {Principles of Optics}}}\ (\bibinfo
  {publisher} {Cambridge Univ. Press},\ \bibinfo {address} {Cambridge},\
  \bibinfo {year} {1999})\BibitemShut {NoStop}%
\bibitem [{\citenamefont {Brosseau}(1998)}]{Brosseau98Book}%
  \BibitemOpen
  \bibfield  {author} {\bibinfo {author} {\bibfnamefont {C.}~\bibnamefont
  {Brosseau}},\ }\href@noop {} {\emph {\bibinfo {title} {Fundamentals of
  Polarized Light}}}\ (\bibinfo  {publisher} {Wiley},\ \bibinfo {address} {New
  York},\ \bibinfo {year} {1998})\BibitemShut {NoStop}%
\bibitem [{\citenamefont {Qian}\ and\ \citenamefont {Eberly}(2011)}]{Qian11OL}%
  \BibitemOpen
  \bibfield  {author} {\bibinfo {author} {\bibfnamefont {X.-F.}\ \bibnamefont
  {Qian}}\ and\ \bibinfo {author} {\bibfnamefont {J.~H.}\ \bibnamefont
  {Eberly}},\ }\href@noop {} {\bibfield  {journal} {\bibinfo  {journal} {Opt.
  Lett.}\ }\textbf {\bibinfo {volume} {36}},\ \bibinfo {pages} {4110} (\bibinfo
  {year} {2011})}\BibitemShut {NoStop}%
\bibitem [{\citenamefont {Kagalwala}\ \emph {et~al.}(2013)\citenamefont
  {Kagalwala}, \citenamefont {{Di G}iuseppe}, \citenamefont {Abouraddy},\ and\
  \citenamefont {Saleh}}]{Kagalwala13NP}%
  \BibitemOpen
  \bibfield  {author} {\bibinfo {author} {\bibfnamefont {K.~H.}\ \bibnamefont
  {Kagalwala}}, \bibinfo {author} {\bibfnamefont {G.}~\bibnamefont {{Di
  G}iuseppe}}, \bibinfo {author} {\bibfnamefont {A.~F.}\ \bibnamefont
  {Abouraddy}}, \ and\ \bibinfo {author} {\bibfnamefont {B.~E.~A.}\
  \bibnamefont {Saleh}},\ }\href@noop {} {\bibfield  {journal} {\bibinfo
  {journal} {Nat. Photon.}\ }\textbf {\bibinfo {volume} {7}},\ \bibinfo {pages}
  {72} (\bibinfo {year} {2013})}\BibitemShut {NoStop}%
\bibitem [{\citenamefont {Abouraddy}\ \emph {et~al.}(2014)\citenamefont
  {Abouraddy}, \citenamefont {Kagalwala},\ and\ \citenamefont
  {Saleh}}]{Abouraddy14OL}%
  \BibitemOpen
  \bibfield  {author} {\bibinfo {author} {\bibfnamefont {A.~F.}\ \bibnamefont
  {Abouraddy}}, \bibinfo {author} {\bibfnamefont {K.~H.}\ \bibnamefont
  {Kagalwala}}, \ and\ \bibinfo {author} {\bibfnamefont {B.~E.~A.}\
  \bibnamefont {Saleh}},\ }\href@noop {} {\bibfield  {journal} {\bibinfo
  {journal} {Opt. Lett.}\ }\textbf {\bibinfo {volume} {39}},\ \bibinfo {pages}
  {2411} (\bibinfo {year} {2014})}\BibitemShut {NoStop}%
\bibitem [{\citenamefont {Kagalwala}\ \emph {et~al.}(2015)\citenamefont
  {Kagalwala}, \citenamefont {Kondakci}, \citenamefont {Abouraddy},\ and\
  \citenamefont {Saleh}}]{Kagalwala15SR}%
  \BibitemOpen
  \bibfield  {author} {\bibinfo {author} {\bibfnamefont {K.~H.}\ \bibnamefont
  {Kagalwala}}, \bibinfo {author} {\bibfnamefont {H.~E.}\ \bibnamefont
  {Kondakci}}, \bibinfo {author} {\bibfnamefont {A.~F.}\ \bibnamefont
  {Abouraddy}}, \ and\ \bibinfo {author} {\bibfnamefont {B.~E.~A.}\
  \bibnamefont {Saleh}},\ }\href@noop {} {\bibfield  {journal} {\bibinfo
  {journal} {Sci. Rep.}\ }\textbf {\bibinfo {volume} {5}},\ \bibinfo {pages}
  {15333} (\bibinfo {year} {2015})}\BibitemShut {NoStop}%
\bibitem [{\citenamefont {Wolf}(2007)}]{Wolf07Book}%
  \BibitemOpen
  \bibfield  {author} {\bibinfo {author} {\bibfnamefont {E.}~\bibnamefont
  {Wolf}},\ }\href@noop {} {\emph {\bibinfo {title} {Introduction to the Theory
  of Coherence and Polarization of Light}}}\ (\bibinfo  {publisher} {Cambridge
  Univ. Press},\ \bibinfo {address} {Cambridge},\ \bibinfo {year}
  {2007})\BibitemShut {NoStop}%
\bibitem [{\citenamefont {Al-Qasimi}\ \emph {et~al.}(2007)\citenamefont
  {Al-Qasimi}, \citenamefont {Korotkova}, \citenamefont {James},\ and\
  \citenamefont {Wolf}}]{Al-Qasimi07OL}%
  \BibitemOpen
  \bibfield  {author} {\bibinfo {author} {\bibfnamefont {A.}~\bibnamefont
  {Al-Qasimi}}, \bibinfo {author} {\bibfnamefont {O.}~\bibnamefont
  {Korotkova}}, \bibinfo {author} {\bibfnamefont {D.}~\bibnamefont {James}}, \
  and\ \bibinfo {author} {\bibfnamefont {E.}~\bibnamefont {Wolf}},\ }\href@noop
  {} {\bibfield  {journal} {\bibinfo  {journal} {Opt. Lett.}\ }\textbf
  {\bibinfo {volume} {32}},\ \bibinfo {pages} {1015} (\bibinfo {year}
  {2007})}\BibitemShut {NoStop}%
\bibitem [{\citenamefont {Abouraddy}(2017)}]{Abouraddy17OE}%
  \BibitemOpen
  \bibfield  {author} {\bibinfo {author} {\bibfnamefont {A.~F.}\ \bibnamefont
  {Abouraddy}},\ }\href@noop {} {\bibfield  {journal} {\bibinfo  {journal}
  {Opt. Express}\ ,\ \bibinfo {pages} {unpublished}} (\bibinfo {year}
  {2017})}\BibitemShut {NoStop}%
\bibitem [{\citenamefont {Brosseau}\ and\ \citenamefont
  {Dogariu}(2006)}]{Brosseau06PO}%
  \BibitemOpen
  \bibfield  {author} {\bibinfo {author} {\bibfnamefont {C.}~\bibnamefont
  {Brosseau}}\ and\ \bibinfo {author} {\bibfnamefont {A.}~\bibnamefont
  {Dogariu}},\ }in\ \href@noop {} {\emph {\bibinfo {booktitle} {Progress in
  Optics Vol. 49}}},\ \bibinfo {editor} {edited by\ \bibinfo {editor}
  {\bibfnamefont {E.}~\bibnamefont {Wolf}}}\ (\bibinfo  {publisher}
  {Elsevier},\ \bibinfo {address} {Amsterdam},\ \bibinfo {year} {2006})\
  Chap.~\bibinfo {chapter} {4}, pp.\ \bibinfo {pages} {315--380}\BibitemShut
  {NoStop}%
\bibitem [{\citenamefont {Gori}\ \emph {et~al.}(2006)\citenamefont {Gori},
  \citenamefont {Santarsiero},\ and\ \citenamefont {Borghi}}]{Gori06OL}%
  \BibitemOpen
  \bibfield  {author} {\bibinfo {author} {\bibfnamefont {F.}~\bibnamefont
  {Gori}}, \bibinfo {author} {\bibfnamefont {M.}~\bibnamefont {Santarsiero}}, \
  and\ \bibinfo {author} {\bibfnamefont {R.}~\bibnamefont {Borghi}},\
  }\href@noop {} {\bibfield  {journal} {\bibinfo  {journal} {Opt. Lett.}\
  }\textbf {\bibinfo {volume} {31}},\ \bibinfo {pages} {858} (\bibinfo {year}
  {2006})}\BibitemShut {NoStop}%
\bibitem [{\citenamefont {Wolf}(2003)}]{Wolf03PLA}%
  \BibitemOpen
  \bibfield  {author} {\bibinfo {author} {\bibfnamefont {E.}~\bibnamefont
  {Wolf}},\ }\href@noop {} {\bibfield  {journal} {\bibinfo  {journal} {Phys.
  Lett. A}\ }\textbf {\bibinfo {volume} {312}},\ \bibinfo {pages} {263}
  (\bibinfo {year} {2003})}\BibitemShut {NoStop}%
\bibitem [{\citenamefont {Tervo}\ \emph {et~al.}(2003)\citenamefont {Tervo},
  \citenamefont {Set{\"a}l{\"a}},\ and\ \citenamefont {Friberg}}]{Tervo03OE}%
  \BibitemOpen
  \bibfield  {author} {\bibinfo {author} {\bibfnamefont {J.}~\bibnamefont
  {Tervo}}, \bibinfo {author} {\bibfnamefont {T.}~\bibnamefont
  {Set{\"a}l{\"a}}}, \ and\ \bibinfo {author} {\bibfnamefont {A.~T.}\
  \bibnamefont {Friberg}},\ }\href@noop {} {\bibfield  {journal} {\bibinfo
  {journal} {Opt. Express}\ }\textbf {\bibinfo {volume} {11}},\ \bibinfo
  {pages} {1137} (\bibinfo {year} {2003})}\BibitemShut {NoStop}%
\bibitem [{\citenamefont {Set{\"a}l{\"a}}\ \emph {et~al.}(2004)\citenamefont
  {Set{\"a}l{\"a}}, \citenamefont {Tervo},\ and\ \citenamefont
  {Friberg}}]{Setala04OE}%
  \BibitemOpen
  \bibfield  {author} {\bibinfo {author} {\bibfnamefont {T.}~\bibnamefont
  {Set{\"a}l{\"a}}}, \bibinfo {author} {\bibfnamefont {J.}~\bibnamefont
  {Tervo}}, \ and\ \bibinfo {author} {\bibfnamefont {A.~T.}\ \bibnamefont
  {Friberg}},\ }\href@noop {} {\bibfield  {journal} {\bibinfo  {journal} {Opt.
  Lett.}\ }\textbf {\bibinfo {volume} {29}},\ \bibinfo {pages} {328} (\bibinfo
  {year} {2004})}\BibitemShut {NoStop}%
\bibitem [{\citenamefont {Gori}\ \emph {et~al.}(2007)\citenamefont {Gori},
  \citenamefont {Santarsiero},\ and\ \citenamefont {Borghi}}]{Gori07OL}%
  \BibitemOpen
  \bibfield  {author} {\bibinfo {author} {\bibfnamefont {F.}~\bibnamefont
  {Gori}}, \bibinfo {author} {\bibfnamefont {M.}~\bibnamefont {Santarsiero}}, \
  and\ \bibinfo {author} {\bibfnamefont {R.}~\bibnamefont {Borghi}},\
  }\href@noop {} {\bibfield  {journal} {\bibinfo  {journal} {Opt. Lett.}\
  }\textbf {\bibinfo {volume} {32}},\ \bibinfo {pages} {588} (\bibinfo {year}
  {2007})}\BibitemShut {NoStop}%
\bibitem [{\citenamefont {Mart{\'i}nez-Herrero}\ and\ \citenamefont
  {Mej{\'i}as}(2007)}]{Herrero07OL}%
  \BibitemOpen
  \bibfield  {author} {\bibinfo {author} {\bibfnamefont {R.}~\bibnamefont
  {Mart{\'i}nez-Herrero}}\ and\ \bibinfo {author} {\bibfnamefont {P.~M.}\
  \bibnamefont {Mej{\'i}as}},\ }\href@noop {} {\bibfield  {journal} {\bibinfo
  {journal} {Opt. Lett.}\ }\textbf {\bibinfo {volume} {32}},\ \bibinfo {pages}
  {1471} (\bibinfo {year} {2007})}\BibitemShut {NoStop}%
\bibitem [{\citenamefont {Peres}(1995)}]{Peres95Book}%
  \BibitemOpen
  \bibfield  {author} {\bibinfo {author} {\bibfnamefont {A.}~\bibnamefont
  {Peres}},\ }\href@noop {} {\emph {\bibinfo {title} {Quantum Theory: Concepts
  and Methods}}}\ (\bibinfo  {publisher} {Kluwer Academic Publishers},\
  \bibinfo {address} {Dordrecht},\ \bibinfo {year} {1995})\BibitemShut
  {NoStop}%
\bibitem [{\citenamefont {Wootters}(1990)}]{Wooters90CEPI}%
  \BibitemOpen
  \bibfield  {author} {\bibinfo {author} {\bibfnamefont {W.~K.}\ \bibnamefont
  {Wootters}},\ }in\ \href@noop {} {\emph {\bibinfo {booktitle} {Complexity,
  Entropy, and the Physics of Information}}},\ \bibinfo {series} {SFI Studies
  in the Sciences of Complexity}, Vol.\ \bibinfo {volume} {VIII},\ \bibinfo
  {editor} {edited by\ \bibinfo {editor} {\bibfnamefont {W.~H.}\ \bibnamefont
  {Zurek}}}\ (\bibinfo  {publisher} {Addison-Wesley},\ \bibinfo {address}
  {Reading},\ \bibinfo {year} {1990})\ pp.\ \bibinfo {pages}
  {39--46}\BibitemShut {NoStop}%
\bibitem [{\citenamefont {James}\ \emph {et~al.}(2001)\citenamefont {James},
  \citenamefont {Kwiat}, \citenamefont {Munro},\ and\ \citenamefont
  {White}}]{James01PRA1}%
  \BibitemOpen
  \bibfield  {author} {\bibinfo {author} {\bibfnamefont {D.~F.~V.}\
  \bibnamefont {James}}, \bibinfo {author} {\bibfnamefont {P.~G.}\ \bibnamefont
  {Kwiat}}, \bibinfo {author} {\bibfnamefont {W.~J.}\ \bibnamefont {Munro}}, \
  and\ \bibinfo {author} {\bibfnamefont {A.~G.}\ \bibnamefont {White}},\
  }\href@noop {} {\bibfield  {journal} {\bibinfo  {journal} {Phys. Rev. A}\
  }\textbf {\bibinfo {volume} {64}},\ \bibinfo {pages} {052312} (\bibinfo
  {year} {2001})}\BibitemShut {NoStop}%
\bibitem [{\citenamefont {Abouraddy}\ \emph {et~al.}(2002)\citenamefont
  {Abouraddy}, \citenamefont {Sergienko}, \citenamefont {Saleh},\ and\
  \citenamefont {Teich}}]{Abouraddy02OptComm}%
  \BibitemOpen
  \bibfield  {author} {\bibinfo {author} {\bibfnamefont {A.~F.}\ \bibnamefont
  {Abouraddy}}, \bibinfo {author} {\bibfnamefont {A.~V.}\ \bibnamefont
  {Sergienko}}, \bibinfo {author} {\bibfnamefont {B.~E.~A.}\ \bibnamefont
  {Saleh}}, \ and\ \bibinfo {author} {\bibfnamefont {M.~C.}\ \bibnamefont
  {Teich}},\ }\href@noop {} {\bibfield  {journal} {\bibinfo  {journal} {Opt.
  Comm.}\ }\textbf {\bibinfo {volume} {201}},\ \bibinfo {pages} {93} (\bibinfo
  {year} {2002})}\BibitemShut {NoStop}%
\bibitem [{\citenamefont {Gori}(1998)}]{Gori98OL}%
  \BibitemOpen
  \bibfield  {author} {\bibinfo {author} {\bibfnamefont {F.}~\bibnamefont
  {Gori}},\ }\href@noop {} {\bibfield  {journal} {\bibinfo  {journal} {Opt.
  Lett.}\ }\textbf {\bibinfo {volume} {23}},\ \bibinfo {pages} {241} (\bibinfo
  {year} {1998})}\BibitemShut {NoStop}%
\bibitem [{\citenamefont {Ellis}\ and\ \citenamefont
  {Dogariu}(2004)}]{Ellis04OL}%
  \BibitemOpen
  \bibfield  {author} {\bibinfo {author} {\bibfnamefont {J.}~\bibnamefont
  {Ellis}}\ and\ \bibinfo {author} {\bibfnamefont {A.}~\bibnamefont
  {Dogariu}},\ }\href@noop {} {\bibfield  {journal} {\bibinfo  {journal} {Opt.
  Lett.}\ }\textbf {\bibinfo {volume} {29}},\ \bibinfo {pages} {536} (\bibinfo
  {year} {2004})}\BibitemShut {NoStop}%
\bibitem [{\citenamefont {Mujat}\ \emph {et~al.}(2004)\citenamefont {Mujat},
  \citenamefont {Dogariu},\ and\ \citenamefont {Wolf}}]{Mujat04JOSAA}%
  \BibitemOpen
  \bibfield  {author} {\bibinfo {author} {\bibfnamefont {M.}~\bibnamefont
  {Mujat}}, \bibinfo {author} {\bibfnamefont {A.}~\bibnamefont {Dogariu}}, \
  and\ \bibinfo {author} {\bibfnamefont {E.}~\bibnamefont {Wolf}},\ }\href@noop
  {} {\bibfield  {journal} {\bibinfo  {journal} {J. Opt. Soc. Am. A}\ }\textbf
  {\bibinfo {volume} {21}},\ \bibinfo {pages} {2414} (\bibinfo {year}
  {2004})}\BibitemShut {NoStop}%
\bibitem [{\citenamefont {Kim}\ \emph {et~al.}(1987)\citenamefont {Kim},
  \citenamefont {Mandel},\ and\ \citenamefont {Wolf}}]{Kim87JOSAA}%
  \BibitemOpen
  \bibfield  {author} {\bibinfo {author} {\bibfnamefont {K.}~\bibnamefont
  {Kim}}, \bibinfo {author} {\bibfnamefont {L.}~\bibnamefont {Mandel}}, \ and\
  \bibinfo {author} {\bibfnamefont {E.}~\bibnamefont {Wolf}},\ }\href@noop {}
  {\bibfield  {journal} {\bibinfo  {journal} {J. Opt. Soc. Am. A}\ }\textbf
  {\bibinfo {volume} {4}},\ \bibinfo {pages} {433} (\bibinfo {year}
  {1987})}\BibitemShut {NoStop}%
\bibitem [{\citenamefont {Gil}(2000)}]{Gil00JOSAA}%
  \BibitemOpen
  \bibfield  {author} {\bibinfo {author} {\bibfnamefont {J.~J.}\ \bibnamefont
  {Gil}},\ }\href@noop {} {\bibfield  {journal} {\bibinfo  {journal} {J. Opt.
  Soc. Am. A}\ }\textbf {\bibinfo {volume} {17}},\ \bibinfo {pages} {328}
  (\bibinfo {year} {2000})}\BibitemShut {NoStop}%
\bibitem [{\citenamefont {Gamel}\ and\ \citenamefont
  {James}(2011)}]{Gamel11OL}%
  \BibitemOpen
  \bibfield  {author} {\bibinfo {author} {\bibfnamefont {O.}~\bibnamefont
  {Gamel}}\ and\ \bibinfo {author} {\bibfnamefont {D.~F.~V.}\ \bibnamefont
  {James}},\ }\href@noop {} {\bibfield  {journal} {\bibinfo  {journal} {Opt.
  Lett.}\ }\textbf {\bibinfo {volume} {36}},\ \bibinfo {pages} {2821} (\bibinfo
  {year} {2011})}\BibitemShut {NoStop}%
\bibitem [{\citenamefont {Simon}\ \emph {et~al.}(2010)\citenamefont {Simon},
  \citenamefont {Simon}, \citenamefont {Gori}, \citenamefont {Santarsiero},
  \citenamefont {Borghi}, \citenamefont {Mukunda},\ and\ \citenamefont
  {Simon}}]{Simon10PRL}%
  \BibitemOpen
  \bibfield  {author} {\bibinfo {author} {\bibfnamefont {B.~N.}\ \bibnamefont
  {Simon}}, \bibinfo {author} {\bibfnamefont {S.}~\bibnamefont {Simon}},
  \bibinfo {author} {\bibfnamefont {F.}~\bibnamefont {Gori}}, \bibinfo {author}
  {\bibfnamefont {M.}~\bibnamefont {Santarsiero}}, \bibinfo {author}
  {\bibfnamefont {R.}~\bibnamefont {Borghi}}, \bibinfo {author} {\bibfnamefont
  {N.}~\bibnamefont {Mukunda}}, \ and\ \bibinfo {author} {\bibfnamefont
  {R.}~\bibnamefont {Simon}},\ }\href@noop {} {\bibfield  {journal} {\bibinfo
  {journal} {Phys. Rev. Lett.}\ }\textbf {\bibinfo {volume} {104}},\ \bibinfo
  {pages} {023901} (\bibinfo {year} {2010})}\BibitemShut {NoStop}%
\bibitem [{\citenamefont {Venkataraman}\ \emph {et~al.}(2013)\citenamefont
  {Venkataraman}, \citenamefont {Saha},\ and\ \citenamefont
  {Gaeta}}]{Gaeta13NP}%
  \BibitemOpen
  \bibfield  {author} {\bibinfo {author} {\bibfnamefont {V.}~\bibnamefont
  {Venkataraman}}, \bibinfo {author} {\bibfnamefont {K.}~\bibnamefont {Saha}},
  \ and\ \bibinfo {author} {\bibfnamefont {A.~L.}\ \bibnamefont {Gaeta}},\
  }\href@noop {} {\bibfield  {journal} {\bibinfo  {journal} {Nat. Photon.}\
  }\textbf {\bibinfo {volume} {7}},\ \bibinfo {pages} {138} (\bibinfo {year}
  {2013})}\BibitemShut {NoStop}%
\bibitem [{\citenamefont {Berg-Johansen}\ \emph {et~al.}(2015)\citenamefont
  {Berg-Johansen}, \citenamefont {T\"oppel}, \citenamefont {Stiller},
  \citenamefont {Banzer}, \citenamefont {Ornigotti}, \citenamefont {Giacobino},
  \citenamefont {Leuchs}, \citenamefont {Aiello},\ and\ \citenamefont
  {Marquardt}}]{Berg15Optica}%
  \BibitemOpen
  \bibfield  {author} {\bibinfo {author} {\bibfnamefont {S.}~\bibnamefont
  {Berg-Johansen}}, \bibinfo {author} {\bibfnamefont {F.}~\bibnamefont
  {T\"oppel}}, \bibinfo {author} {\bibfnamefont {B.}~\bibnamefont {Stiller}},
  \bibinfo {author} {\bibfnamefont {P.}~\bibnamefont {Banzer}}, \bibinfo
  {author} {\bibfnamefont {M.}~\bibnamefont {Ornigotti}}, \bibinfo {author}
  {\bibfnamefont {E.}~\bibnamefont {Giacobino}}, \bibinfo {author}
  {\bibfnamefont {G.}~\bibnamefont {Leuchs}}, \bibinfo {author} {\bibfnamefont
  {A.}~\bibnamefont {Aiello}}, \ and\ \bibinfo {author} {\bibfnamefont
  {C.}~\bibnamefont {Marquardt}},\ }\href@noop {} {\bibfield  {journal}
  {\bibinfo  {journal} {Optica}\ }\textbf {\bibinfo {volume} {2}},\ \bibinfo
  {pages} {864} (\bibinfo {year} {2015})}\BibitemShut {NoStop}%
\bibitem [{\citenamefont {T\"oppel}\ \emph {et~al.}(2014)\citenamefont
  {T\"oppel}, \citenamefont {Aiello}, \citenamefont {Marquardt}, \citenamefont
  {Giacobino},\ and\ \citenamefont {Leuchs}}]{Toppel14NJP}%
  \BibitemOpen
  \bibfield  {author} {\bibinfo {author} {\bibfnamefont {F.}~\bibnamefont
  {T\"oppel}}, \bibinfo {author} {\bibfnamefont {A.}~\bibnamefont {Aiello}},
  \bibinfo {author} {\bibfnamefont {C.}~\bibnamefont {Marquardt}}, \bibinfo
  {author} {\bibfnamefont {E.}~\bibnamefont {Giacobino}}, \ and\ \bibinfo
  {author} {\bibfnamefont {G.}~\bibnamefont {Leuchs}},\ }\href@noop {}
  {\bibfield  {journal} {\bibinfo  {journal} {New J. Phys.}\ }\textbf {\bibinfo
  {volume} {16}},\ \bibinfo {pages} {073019} (\bibinfo {year}
  {2014})}\BibitemShut {NoStop}%
\bibitem [{\citenamefont {Tripathi}\ and\ \citenamefont
  {Toussaint}(2009)}]{Tripathi09OE}%
  \BibitemOpen
  \bibfield  {author} {\bibinfo {author} {\bibfnamefont {S.}~\bibnamefont
  {Tripathi}}\ and\ \bibinfo {author} {\bibfnamefont {K.~C.}\ \bibnamefont
  {Toussaint}},\ }\href@noop {} {\bibfield  {journal} {\bibinfo  {journal}
  {Opt. Express}\ }\textbf {\bibinfo {volume} {17}},\ \bibinfo {pages} {21396}
  (\bibinfo {year} {2009})}\BibitemShut {NoStop}%
\bibitem [{\citenamefont {Kondakci}\ and\ \citenamefont
  {Abouraddy}(2016)}]{Kondakci16OE}%
  \BibitemOpen
  \bibfield  {author} {\bibinfo {author} {\bibfnamefont {H.~E.}\ \bibnamefont
  {Kondakci}}\ and\ \bibinfo {author} {\bibfnamefont {A.~F.}\ \bibnamefont
  {Abouraddy}},\ }\href@noop {} {\bibfield  {journal} {\bibinfo  {journal}
  {Opt. Express}\ }\textbf {\bibinfo {volume} {24}},\ \bibinfo {pages} {28659}
  (\bibinfo {year} {2016})}\BibitemShut {NoStop}%
\bibitem [{\citenamefont {Parker}\ and\ \citenamefont
  {Alonso}(2016)}]{Parker16OE}%
  \BibitemOpen
  \bibfield  {author} {\bibinfo {author} {\bibfnamefont {K.~J.}\ \bibnamefont
  {Parker}}\ and\ \bibinfo {author} {\bibfnamefont {M.~A.}\ \bibnamefont
  {Alonso}},\ }\href@noop {} {\bibfield  {journal} {\bibinfo  {journal} {Opt.
  Express}\ }\textbf {\bibinfo {volume} {24}},\ \bibinfo {pages} {28669}
  (\bibinfo {year} {2016})}\BibitemShut {NoStop}%
\end{thebibliography}%
\end{document}